\documentclass[letterpaper, preprint, paper,11pt]{AAS}	

\usepackage{bm}
\usepackage{amsmath}
\usepackage[colorlinks=true, pdfstartview=FitV, linkcolor=black, citecolor= black, urlcolor= black]{hyperref}
\usepackage{overcite}
\usepackage{footnpag}			      	

\usepackage[font=scriptsize]{caption}
\usepackage[font=scriptsize]{subcaption} 

\PaperNumber{24-493}

\begin{document}

\title{Meshing of High-Dimensional Toroidal Manifolds from Quasi-Periodic Three-Body Problem Dynamics using Parameterization via Discrete One-Forms}

\author{Dante Basile\thanks{PhD Student, Department of Computer Science, Purdue University, 610 Purdue Mall, West Lafayette, IN 47907, dbasile@purdue.edu},  
Xavier Tricoche\thanks{Associate Professor, Department of Computer Science, Purdue University, 610 Purdue Mall, West Lafayette, IN 47907, xmt@purdue.edu},
\ and Martin Lo\thanks{Principal Engineer, Mission Design and Navigation Section, Jet Propulsion Laboratory, California Institute of Technology, Pasadena, CA 91109, United States, martin.w.lo@jpl.nasa.gov}
}

\maketitle{}

\begin{abstract}
High-dimensional visual computer models are poised to revolutionize the space mission design process. The circular restricted three-body problem (CR3BP) gives rise to high-dimensional toroidal manifolds that are of immense interest to mission designers. We present a meshing technique which leverages an embedding-agnostic parameterization to enable topologically accurate modelling and intuitive visualization of toroidal manifolds in arbitrarily high-dimensional embedding spaces. This work describes the extension of a discrete one-form-based toroidal point cloud meshing method to high-dimensional point clouds sampled along quasi-periodic orbital trajectories in the CR3BP. The resulting meshes are enhanced through the application of an embedding-agnostic triangle-sidedness assignment algorithm. This significantly increases the intuitiveness of interpreting the meshes after they are downprojected to 3D for visualization. These models provide novel surface-based representations of high-dimensional topologies which have so far only been shown as points or curves. This success demonstrates the effectiveness of differential geometric methods for characterizing manifolds with complex, high-dimensional embedding spaces, laying the foundation for new models and visualizations of high-dimensional solution spaces for dynamical systems. Such representations promise to enhance the utility of the three-body problem for the visual inspection and design of space mission trajectories by enabling the application of proven computational surface visualization and analysis methods to underlying solution manifolds.
\end{abstract}

\section{Introduction}

In this work, we develop a point cloud meshing technique for the modelling and visualization of solution spaces for the circular restricted three-body problem (CR3BP). Spacecraft position and motion in the CR3BP are described by a dynamical system~\cite{regchaodynam}. Leveraging the dynamics of the CR3BP for space mission trajectory design is essential because reducing fuel consumption increases scientific payload capacity. Ballistic trajectories, or trajectories on which the spacecraft is influenced by gravity alone, do not require fuel consumption. The dynamics of the CR3BP predict ballistic trajectories along toroidal (donut shaped) manifolds. These dynamics are studied in a phase space where both 3D position and velocity are needed to define all parameters relevant to trajectory design. Thus, the manifolds exist in 4D-6D space, making it challenging to represent them with insightful visualizations. To address this challenge, we present a parameterization-based embedding-agnostic meshing method for high-dimensional toroidal point clouds. This approach promises to increase the effectiveness and intuitiveness of space mission design by enabling mission designers to model and visualize numerically sampled point cloud solution spaces with their true, and often never-before-seen, surface representation.

\begin{figure}[htb]
    \centering
    \includegraphics[width=.5\textwidth]{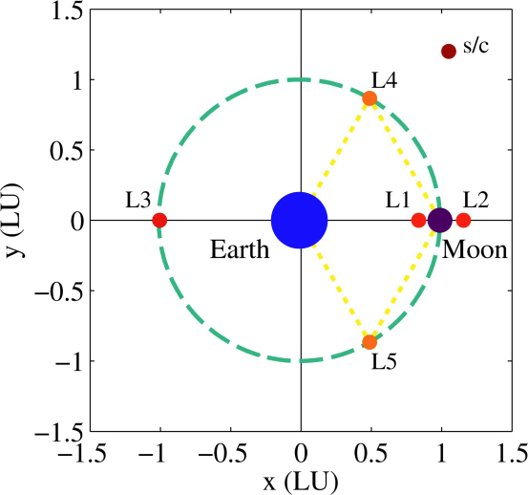}
    \caption{The CR3BP is shown in the rotating frame. The Earth is $m_1$, represented by the blue dot, and the Moon is $m_2$, represented by the purple dot. $m_1$ and $m_2$ are located on the x-axis at 0 and 1 respectively. The dark red dot labeled s/c represents the spacecraft, $m_3$, on its approach. The libration points L1, L2, and L3 are represented by red dots. The libration points L4 and L5 are represented by orange dots. The green dashed line shows the orbit of $m_2$ about $m_1$. The yellow dotted lines illustrate the location of L4 and L5 at the third vertex of the equilateral triangles formed by $m_1$, $m_2$, and either libration point.}
    \label{fig:cr3bpp}
\end{figure}

The CR3BP is a dynamical system describing the position and velocity of three bodies which orbit each other in space~\cite{mathcelmech, r3bp}. We consider the case where a massless body $m_3$, often a spacecraft, orbits two massive bodies $m_1$ and $m_2$ representing the sun and a planet or a planet and a moon. We apply standard barycentric normalized coordinates, defining the mass parameter $\mu$ as
\begin{equation}
    \mu = \frac{m_2}{m_1 + m_2}
\end{equation}
such that $m_1$ is the mass of the primary body and $m_2$ is the mass of the secondary body. According to convention, $m_1 \ge m_2$. This results in the constraint $\mu \in (0, 0.5)$.

$m_1$ and $m_2$ exist in a circular orbit about their barycenter. We consider the CR3BP in the rotating frame such that $m_1$ and $m_2$ are fixed on the x-axis at $x_1 = -\mu$ and $x_2 = 1 - \mu$. A visual representation of this arrangement is provided in Figure~\ref{fig:cr3bpp}. The state $\mathbf{x}$ of $m_3$ is defined as
\begin{align}
    \mathbf{r} &= [x \quad y \quad z]^\mathsf{T}\\
    \mathbf{v} &= [\dot{x} \quad \dot{y} \quad \dot{z}]^\mathsf{T}\\
    \mathbf{x} &=
    \begin{bmatrix}
        \mathbf{r}\\
        \mathbf{v}
    \end{bmatrix}
\end{align}
where $\mathbf{r}$ represents the position and $\mathbf{v}$ represents the velocity.

The mass of $m_3$ is considered to be negligible relative to that of $m_1$ and $m_2$. Therefore, we restrict it to have no gravitational effect on either of the larger bodies, specifying the \textit{restricted} three-body problem. The equations of motion in the CR3BP rotating frame are thus
\begin{align}
    \Ddot{x} &= -\biggl((1-\mu) \frac{x + \mu}{\|\mathbf{r}_1\|^3} + \mu \frac{x - 1 + \mu}{\|\mathbf{r}_2\|^3}\biggr) + x + 2 \dot{y}\\
    \Ddot{y} &= -\biggl((1-\mu) \frac{y}{\|\mathbf{r}_1\|^3} + \mu \frac{y}{\|\mathbf{r}_2\|^3}\biggr) + y - 2 \dot{x}\\
    \Ddot{z} &= -\biggl((1-\mu) \frac{z}{\|\mathbf{r}_1\|^3} + \mu \frac{z}{\|\mathbf{r}_2\|^3}\biggr)
\end{align}
where $\mathbf{r}_1$ and $\mathbf{r}_2$ are defined as
\begin{align}
    \mathbf{r}_1 &= [(x + \mu),~y,~z]^\mathsf{T}\\
    \mathbf{r}_2 &= [(x - 1 + \mu),~y,~z]^\mathsf{T}
\end{align}
The augmented potential equation is
\begin{equation}
    U(\mathbf{r}) = \frac{(x^2 + y^2)}{2} + \frac{(1 - \mu)}{\|\mathbf{r}_1\|} = \frac{\mu}{\|\mathbf{r}_2\|}
\end{equation}
which defines the Jacobi constant as
\begin{equation}
    C(\mathbf{r}, \mathbf{v}) = 2 U(\mathbf{r}) - \mathbf{v}^\mathsf{T}\mathbf{v}
\end{equation}

The five equilibrium points of the CR3BP are known as libration points or Lagrange points and are denoted as L1-L5. The Jacobi constant of a libration point is computed by setting $\mathbf{v} = \mathbf{0}$.

Mission designers rely on their understanding of the CR3BP to identify specific orbits which offer explicit, long-term ballistic solutions that maintain proximity with or repeatedly fly by objects of interest. Many orbits possess periodic or quasi-periodic behavior that can be exploited to reduce the complexity of the solution space. Periodic orbits follow an identical path each period and are confined to a closed curve, or 1-torus. Quasi-periodic orbits (QPOs) have a path that evolves with each period and are confined to a donut, or 2-torus~\cite{regchaodynam}. This torus is an invariant manifold, meaning that orbital trajectories are constrained to its surface by the constant energy of the system.  Quasi-periodic orbits are abundant in phase space. They provide mission designers with a broad range of trajectories ensuring design flexibility. Additionally, QPOs allow a spacecraft to cover a wide area over time and can be used for observation of a general region, such as a moon’s entire surface.

Currently, the manifolds are represented by point clouds sampled from numerical integration of the dynamical system. Periodic phenomena occurring along manifolds can be captured using Poincar\'e maps~\cite{poincelestmechdschao, vistopostrucareapresmap}. These models have demonstrated the effectiveness of using visual techniques to guide space mission trajectory design~\cite{levquasperorbtrajdescislun, trajdessatoceaworlorbmdpoinmap}. However, this method is fundamentally limited in terms of the number of state-space dimensions it can interpret and thus is restricted to the 2D Planar CR3BP (PCR3BP) in which the 4D phase space can be reduced to a 2D Poincar\'e map using both a simplification based on the Jacobi constant and restriction to the Poincar\'e section~\cite{extrvispoinspacecrafttrajdes}. Considering motion outside of the equatorial plane in the CR3BP adds an additional position and velocity to the phase space and yields 4D Poincar\'e maps that are nontrivial to visualize. Considering the Elliptic Restricted 3 Body Problem (ER3BP) prevents the Jacobi constant simplification in the phase space formulation, further increasing the dimensionality by one. As analysis seeks to move beyond the PCR3BP towards the full 3D case with a 6D phase space, surface representations compatible with high-dimensional embedding spaces will be required. Therefore, expanding the model’s dimensionality is of critical benefit as missions seek increasingly close and comprehensive flybys~\cite{NRHOfunnels}.

Advances in high-dimensional modelling and visualization are increasingly in-demand at the cutting edge of space mission design. Missions are reaching levels of complexity in which previously valid assumptions begin to break down. For example, missions such as Europa Clipper involve close flybys of gas giant moons. In the case of Europa Clipper, close flybys are necessary for detailed study of the surface in order to detect landing sites for future missions and evaluate the possibility of a subsurface liquid ocean. Follow-on missions to Europa Clipper like the Europa Lander concept will require the design of both ballistic capture and landing maneuvers for Europa. Such missions will need to leverage QPOs to achieve trajectories of increasing complexity. Motion outside the equatorial plane is necessary to study the entire surface of a body. In particular, there are many bodies of which the polar regions are of specific interest. Polar regions are difficult to study from the equatorial plane, causing models that extend beyond the planar assumption to be advantageous for their study. Planning for such missions would be greatly enhanced by visual models that can represent the full 3D CR3BP. Additionally, Lissajous and Halo orbits are prominent examples of trajectories that extend beyond the equatorial plane. These orbits have been of increasing interest for lunar missions. Elliptical orbits allow close flybys of body surfaces and are additionally useful for situations where a mission desires flybys of a body other than the body orbited by the spacecraft. Both of these scenarios are relevant to the upcoming Europa Clipper mission. Close flybys are necessary to scout the surface for suitable landing sites. Moreover, the Clipper spacecraft will orbit Jupiter, with elliptical trajectories allowing it to travel out to Europa's location periodically.

To address this multifaceted need for increasingly complex models reliant on fewer assumptions, models can leverage high-dimensional surface meshes. These allow discrete numerical solutions in the form of point clouds to approximate a continuous surface formed of triangles. Approximations are essential because dense sampling by integrating trajectories along the invariant manifolds is not computationally tractable. However, downprojection of high-dimensional surfaces can lead to apparent self-intersections. The toroidal manifolds of the QPOs are topologically non-trivially embedded in the energy surface. Because of this, projecting the manifolds to 3D for visualization leads to self-intersection artifacts. The embedded torus in configuration space turns inside-out twice to preserve its orientability. Visualizing the stable and unstable manifold of this torus presents great challenges. Furthermore, these tori have a complex fractal hierarchy in phase space~\cite{tori4Dsymp}. The spatial complexity of such phenomena motivates the use of meshes native to the high-dimensional embedding space along with the visualization techniques required to make their interpretation intuitive. The standard method for triangulating a field of points is Delaunay triangulation. However, Delaunay triangulation only generates 2-dimensional surface manifolds when applied to point clouds that exist in a 2-dimensional embedding space. Therefore, a 2D parameterization for the point cloud is required. If this low-dimensional representation is topologically valid, we can simply apply the triangulation results to mesh the original point cloud. This allows Delaunay triangulation to be leveraged to mesh 2-dimensional surfaces that are immersed in higher-dimensional space.

\section{Methods}

\begin{figure}[hp]
    \centering
    \begin{subfigure}{\textwidth}
        \centering
        \includegraphics[width=.9\textwidth]{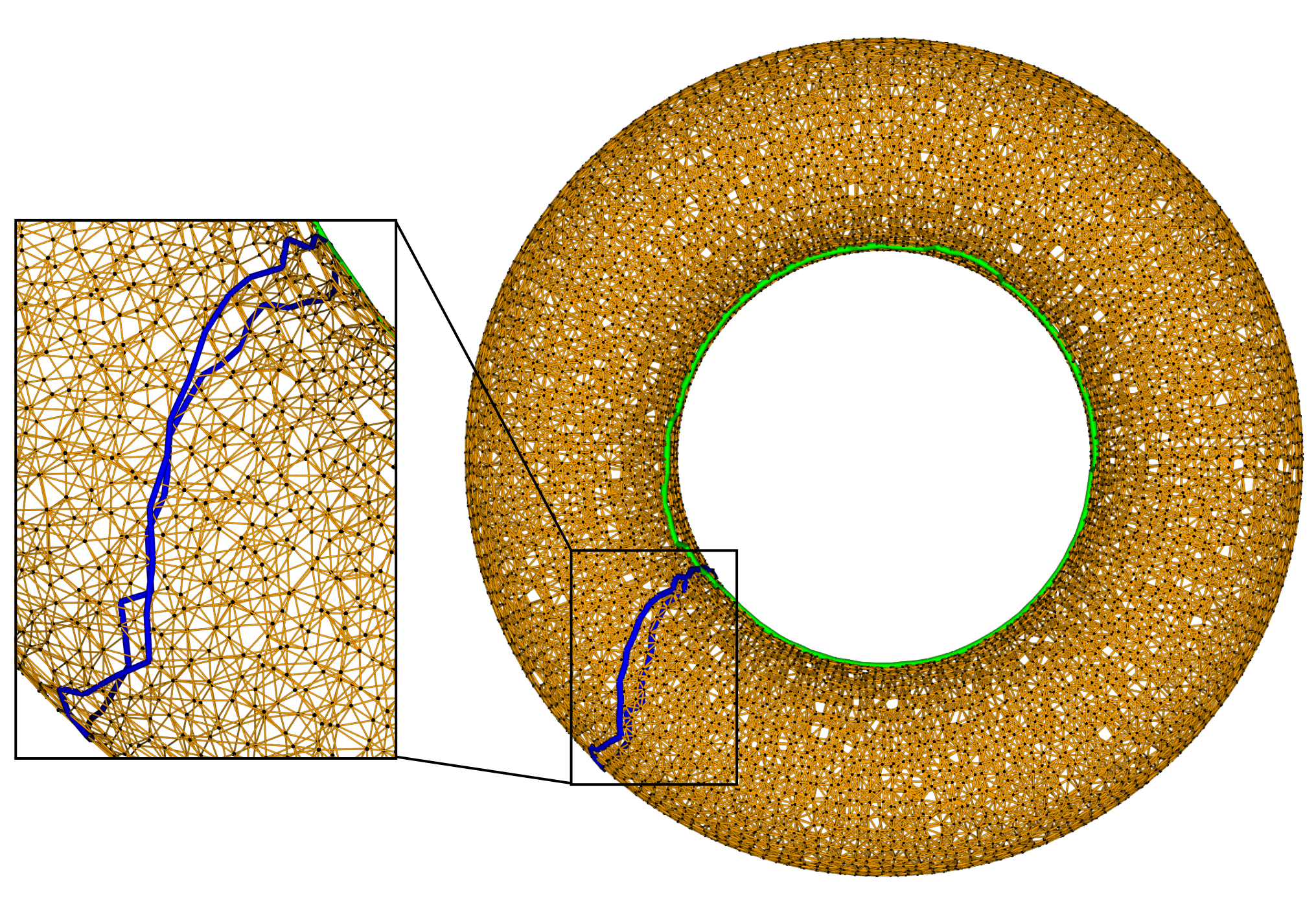}
        \caption{3D toroidal manifold from synthetic data}
    \end{subfigure}
    \begin{subfigure}{\textwidth}
        \centering
        \includegraphics[width=.8\textwidth]{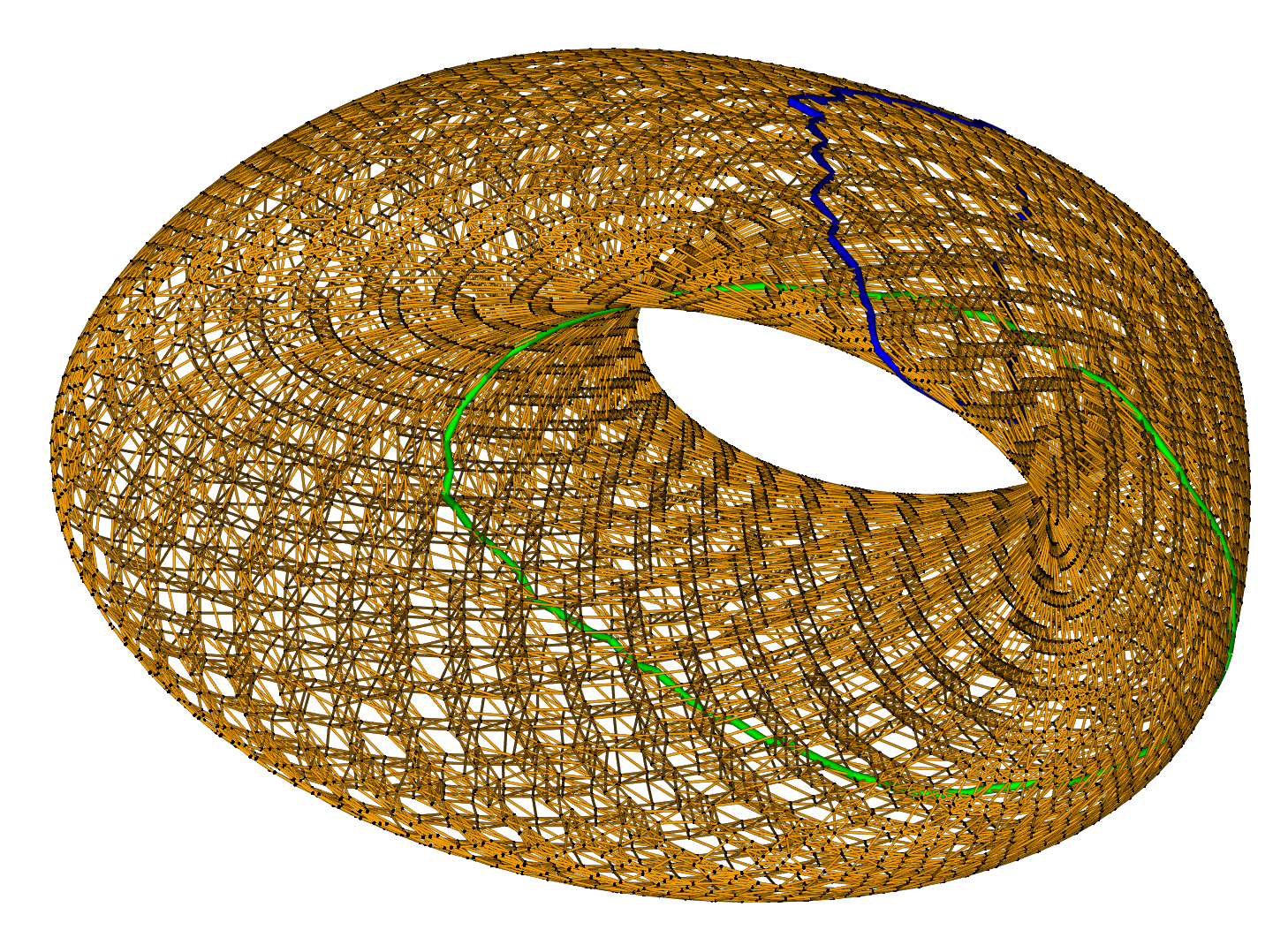}
        \caption{4D toroidal manifold sampled by standard map, downprojected to 3D}
    \end{subfigure}
    \caption{In a KNNG with toroidal topology (edges shown in orange), the two largest cycles in the MCB are considered the non-trivial cycles. They capture the periodicity of a toroidal manifold along both the toroidal and poloidal axes. The toroidal cycle (edges shown in green) encircles the empty space at the center of the torus, taking the long way to return to its start vertex. The poloidal cycle (edges shown in blue) encircles the surface of the torus itself, taking the short way to return to its start vertex. The numerically computed point set is shown in black.}
    \label{fig:mcbntc}
\end{figure}

Our high-dimensional meshing technique is a novel extension of a parameterization-based method originally designed for the meshing of 2-tori in 3D by Tewari et al. \cite{meshgenus1}. We exploit the knowledge that the point cloud is sampled from a 2-manifold with toroidal topology to create a parameterization that explicitly enforces topological correctness. First, point connectivity is established using a k-nearest neighbors graph (KNNG). Then, the minimum cycle basis (MCB) of this graph is computed \cite{mcbalgos}. The topology of a torus guarantees that the two largest cycles in this basis capture the two axes of periodicity while the rest of the cycles represent closed boundary loops. This is because a torus exhibits periodic behavior with respect to both its toroidal and poloidal coordinates. An example of this behavior is demonstrated in Figure~\ref{fig:mcbntc}. This information can be used to construct the necessary co-closedness and closedness equations to consider the edges of the KNNG graph as discrete one-forms.

The description of this approach given in Tewari et al. is included here for completeness. 2D mesh parameterization methods typically flatten meshes to the plane such that the resulting faces are disjoint and minimally distorted. Tutte~\cite{h2dgraph} showed that a manifold mesh with disk topology can be parameterized such that each interior vertex is positioned in the plane at the centroid of its neighbors' positions. Tutte then proved for a planar graph $G$ with $B$ boundary vertices, $V$ interior vertices, $E$ edges, and $F$ faces that if $G$ is 3-connected, as in a triangulation, then the faces in the plane have positive area and are disjoint. This embedding minimizes the sum of the squares of the edge lengths among all drawings of $G$ in the plane with the same boundary conditions. Furthermore, Floater~\cite{paramapproxsurftri} proved that such an embedding still exists for weighted edges, given that each interior vertex $v$ is positioned at some arbitrary convex combination of the positions of its neighbors $x_i$
\begin{equation}
    \forall i \in \{1, \ldots, V\} \quad x_i = \sum_{j \in N(i)} {w_{i j} x_j} \quad \sum_{j \in N(i)} {w_{i j} = 1}, \quad \forall j \quad w_{i j} \ge 0
\end{equation}
where $N(i)$ is the set of neighbors to vertex $i$. This can be algebraically simplified to
\begin{equation}
    \forall i \in \{1, \ldots, V\} \quad 0 = \sum_{j \in N(i)} {w_{i j} (x_i - x_j)}
    \label{eq:d1fs}
\end{equation}
Discrete one-forms are half-edges representing pairwise distances between sample points. They are defined as $\Delta x_{i j} = x_i - x_j$, where $\Delta x_{i j} = \Delta x_{j i}$. Using discrete one-forms and Eq.~\eqref{eq:d1fs}, we define the co-closedness equations to ensure that the value at each vertex is the convex combination of its neighbors
\begin{equation}
    \forall v \in \{1, \ldots, V\} \quad 0 = \sum_{e \in \delta v} {w_e \Delta x_e}
    \label{eq:cce}
\end{equation}
where $v$ is a vertex, $e$ is a half-edge, and $\delta v$ is the set of half-edges defined as directed away from $v$. The closedness equations ensure that cycles return to their initial value, causing the discrete one-forms to fit together as face boundaries
\begin{equation}
    \forall f \in \{1, \ldots, F\} \quad 0 = \sum_{e \in \partial f} {\Delta x_e}
    \label{eq:ce}
\end{equation}
where $f$ is a face and $\partial f$ is the set of half-edges bounding $f$. The method considers faces as cycles. Recall that all but the two largest cycles of the MCB of a torus represent closed boundary loops. We can consider these closed boundary loops as equivalent to $\partial f$. Then, Eq.~\eqref{eq:ce} is reformulated as
\begin{equation}
    \forall c \in \{1, \ldots, C-2\} \quad 0 = \sum_{e \in \partial f} {\Delta x_e}
    \label{eq:cemcb}
\end{equation}
where $\{1, \ldots, C-2\}$ represent all but the last two cycles in a list of the cycles in the MCB ordered from least to greatest length, equivalently the cycles that represent closed boundary loops by the previous definition. Together, Eq.~\eqref{eq:cce} and Eq.~\eqref{eq:cemcb} specify a parameterization that is integrable within any closed patch on the surface. Solving this system of equations, we obtain a 2D parameterization in which the manifold surface can be Delaunay triangulated. The mesh is then constructed in the complete set of dimensions from patches that are triangulated in parameter space. These patches are taken from the torus using breadth-first search to expand out from randomly sampled seed vertices on the KNNG. The topology of the torus guarantees that these patches will have disk topology. Therefore, they are compatible with the aforementioned parameterization technique. The rigorous construction of this parameterization guarantees both agreement between patches and topological correctness given sufficient sampling density.

We have advanced this method, enabling it to function on 2-tori immersed in 4D-6D space. Construction of the parameterization requires only the discrete one-forms to integrate along the manifold surface. These are defined using the co-closedness and closedness equations. In turn, this system of equations solely depends on the distance function used to establish connectivity via the KNNG and the guarantees that toroidal topology is enforced with respect to the MCB. Both of these criteria are agnostic to the dimensionality of the embedding space, allowing this method to function in embedding spaces of arbitrarily high dimensions. Unlike existing techniques, this algorithm provides both compatibility with high-dimensional embedding spaces and topological guarantees to represent the double periodicity of the torus.

The manifold is meshed and then downprojected to a three-dimensional embedding space for interactive visualization. The mesh representation allows us to assign and color manifold sidedness based solely on triangle adjacency. Sidedness is defined for a seed triangle by assigning an arbitrary order to its vertices. This ordering defines a direction on the triangle's edges. Orientation is then propagated along the mesh by assigning ordering to neighboring triangles such that each pair of neighbors sharing an edge assign opposite directions to the edge they share. Triangle vertex orders implicitly define orientation in a manner analogous to the cross product. Assigning orientation solely based on vertex order allows the dimensionality of the embedding space to be successfully disregarded. This results in a visual representation of the sidedness of the manifold in its high-dimensional native space that is seamlessly integrated with the intuitive 3D surface representation. Because this assignment is accurate with respect to the high-dimensional space, it highlights self-intersection artifacts caused by the downprojection to 3D. The computational performance of the parameterization, meshing process, and visualizations are high due to their ability to exploit parallelization. Triangle meshes are fundamental graphics primitives and can leverage GPU compute to meet the requirements for real-time user interaction.

\section{Results}

\begin{figure}[hp]
    \centering
    \begin{subfigure}{\textwidth}
        \centering
        \includegraphics[width=.75\textwidth]{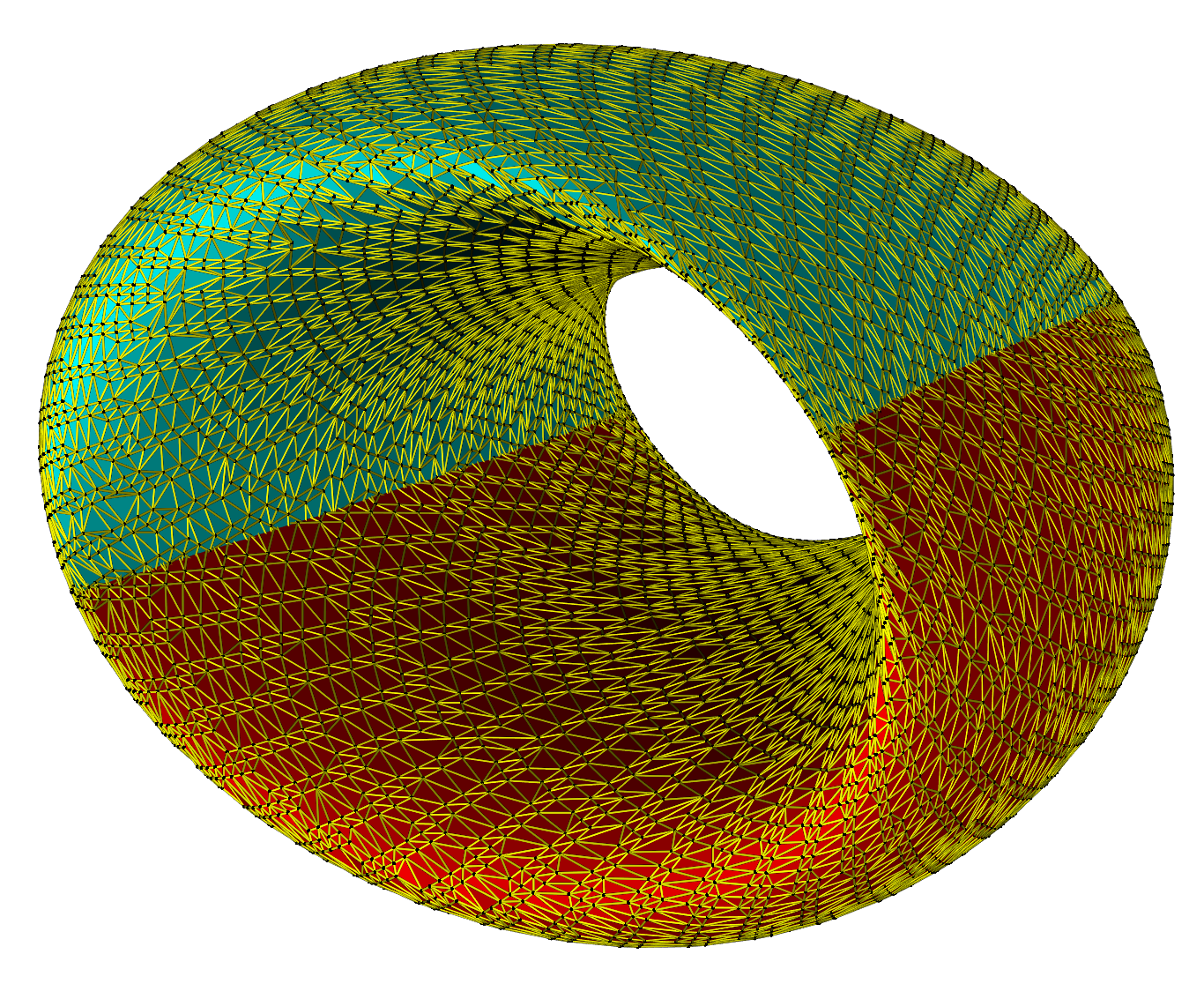}
        \caption{4D toroidal manifold sampled by standard map, downprojected to 3D}
        \label{fig:meshsidea}
    \end{subfigure}
    \begin{subfigure}{\textwidth}
        \centering
        \includegraphics[width=.85\textwidth]{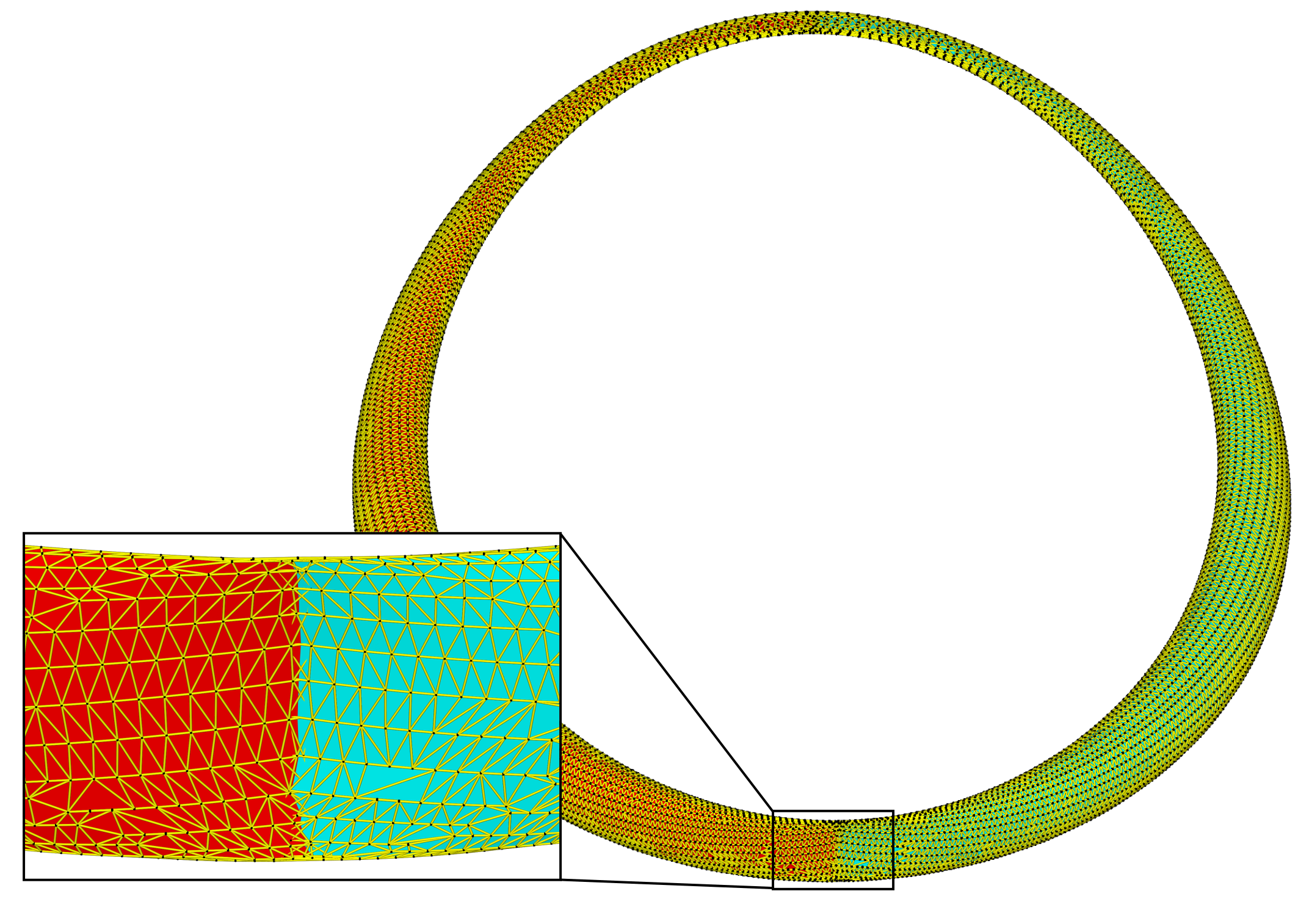}
        \caption{6D toroidal manifold sampled by numerical integration of quasi-periodic orbital trajectories in the Saturn-Enceladus system, downprojected to 3D}
        \label{fig:meshsideb}
    \end{subfigure}
    \caption{The high-dimensional manifolds are meshed in their native embedding space using the parameterization. The resulting meshes are down-projected to 3D for visualization. Triangle edges are shown in yellow. Triangle color indicates mesh sidedness to aid interpretation of the self-intersection artifacts resulting from the down-projection. The numerically computed point set is shown in black.}
    \label{fig:meshside}
\end{figure}

Our work has produced the first known mesh of a 4D manifold sampled with a standard map. This exciting result is shown in Figure~\ref{fig:meshsidea}. Because the point cloud is generated using a map, there is no high-level trajectory structure to exploit, emphasizing the importance of our embedding-agnostic, parameterization-based approach. Additionally, our work has produced the first successful meshing of an invariant manifold embedded in the 6D state-space of orbital trajectory solutions for the south pole of Saturn’s moon, Enceladus, shown in Figure~\ref{fig:meshsideb}. This data was provided by a team of mission designers at NASA JPL who have shown great interest in our technique and who are eager to use this code in their future work. Accurate representations of high-dimensional manifolds will be of great utility to future missions attempting close flybys to search for signs of life in the liquid ejections from the south pole.

\section{Discussion}

The surface representation produced by the mesh enables the use of a wide range of techniques that were not applicable to the previous point cloud representation. Examples include inter-point connectivity, traversal along the surface, sidedness identification, boundary detection, intersection or collision detection, continuous interpolation of new points on the surface between known solutions, and topological representation and analysis of the manifold. One avenue for future work is using surface-based interpolation to sample regions of interest in the state-space. Such an approach is much more computationally efficient than locating and integrating a trajectory that traverses such regions, enhancing the effectiveness of real-time user interaction.

Further opportunity for model improvement lies in the potential for mesh optimization. There is significant existing literature on mesh refinement methods. However, much of this was developed for lower-dimensional models with increased mesh flexibility such as LIDAR data in positional dimensions. In contrast, our model considers structures that exist beyond visual space. Additionally, our point set is incompatible with deformability assumptions because it must maintain its validity as a representation of very accurate numerical solutions to the dynamical system. Considering this, moving points is difficult to justify in our case. Similarly, adding points to sample sparse regions is difficult because a suitable trajectory must first be located and integrated. Therefore, this modelling technique stands to benefit from the development of specialized mesh refinement techniques for high-dimensional dynamical systems. Suitable mesh improvements must center on edge selection. The corresponding modifications will be built directly inside the meshing algorithm.

There is also opportunity for expanding the scope of the manifolds to which this method can be applied. Currently, its parameterization is reliant on the topological properties of the MCB of a torus, restricting its application to toroidal manifolds. However, discrete one-forms remain the core primitive of this method. One-forms represent a fundamental characteristic of all surface manifolds, and discrete one-forms are likewise applicable to any graph induced on their topology. Accordingly, there exists the potential to leverage both established and as of yet undeveloped alternative parameterization schemes suitable for a greater variety of manifold topologies. Furthermore, topological methods which are not directly based on Delaunay triangulation raise the potential of meshing techniques which possess drastically different approaches to parameterization or lack a parameterization altogether.

Additionally, our method greatly enhances the intuitiveness of visualizations for high-dimensional manifolds. Sidedness visualization provides much needed clarity with respect to the self-intersection artifacts caused by downprojecting high-dimensional manifolds to 3D space. Further work on visualization is required because downprojection remains an imperfect solution, especially in 5D and 6D spaces where multiple dimensions must be disregarded simultaneously. Conventional computer screens only directly represent 2D spaces, and human spatial reasoning only applies to 3D. Therefore, dimensionality reduction is needed to extend visualization capability to the desired 4D-6D. Previous efforts have demonstrated the potential for advancement in this direction, and this is an active area of our current investigation. ~\cite{tori4Dsymp, highdimpoinsctraj}.

\section{Conclusion}

This method has generated unprecedented computational surface representations of toroidal manifolds embedded in higher dimensions for both theoretical and astrodynamical applications. However, there exist several promising opportunities for improving the quality, versatility, and presentation of this method's results. This will expand the scope of its application to an increasingly wider variety of dynamical modelling challenges throughout the domain of space mission design and beyond. Ultimately, this method demonstrates the untapped potential of computer graphics and differential geometric methods to provide accurate and intuitive visualizations of high-dimensional orbital dynamics for space mission design.

\section{Acknowledgements}

This work was carried out in part at Purdue University and in part at the Jet Propulsion Laboratory, California Institute of Technology, under a contract with the National Aeronautics and Space Administration (80NM0018D0004).

\bibliographystyle{AAS_publication}   
\bibliography{references}   

\end{document}